\documentclass[Afour,sageh,times]{sagej}

\usepackage{graphicx}
\graphicspath{{./figures/}}
\usepackage[T1]{fontenc}
\usepackage{txfonts}
\usepackage{mathptmx}
\usepackage{booktabs}
\usepackage{hyperref}
\usepackage{soul}
\usepackage{tabularx}
\newcolumntype{C}{>{\centering\arraybackslash}X}
\newcommand{\declaration}[1]{\subsubsection*{\noindent\normalsize\sagesf\bfseries #1:}}

\setcitestyle{numbers}

\def\volumeyear{2023}
\def\journalname{Digital Health}

\begin{document}

\runninghead{Mendu et al.}

\title{Designing Voice Interfaces to Support Mindfulness-Based Pain Management}

\author{Sanjana Mendu\affilnum{1}, Sebrina L. Doyle Fosco\affilnum{2}, Stephanie T. Lanza\affilnum{2}, and Saeed Abdullah\affilnum{1}}

\affiliation{\affilnum{1}College of Information Sciences and Technology, Pennsylvania State University \\
\affilnum{2}Edna Bennett Pierce Prevention Research Center, Pennsylvania State University}

\corrauth{Sanjana Mendu
College of Information Sciences and Technology,
Pennsylvania State University,
Westgate Building,
University Park, PA 16802}
\email{spm6450@psu.edu}
\twitter{sanjana\_mendu}

\begin{abstract}
\textbf{Objective:} Chronic pain is a critical public health issue affecting approximately 20\% of the adult population in the United States. Given the opioid crisis, there has been an urgent focus on non-addictive pain management methods including Mindfulness-Based Stress Reduction (MBSR). Prior work has successfully used MBSR for pain management. However, ensuring longitudinal engagement to MBSR practices remains a serious challenge. In this work, we explore the utility of a voice interface to support MBSR home practice.

\textbf{Methods:} We interviewed ten mindfulness program facilitators to understand how such a technology might fit in the context of the MBSR class and identify potential usability issues with our prototype. We then used directed content analysis to identify key themes and sub-themes within the interview data.

\textbf{Results:} Our findings show that facilitators supported the use of the voice interface for MBSR, particularly for individuals with limited motor function. Facilitators also highlighted unique affordances of voice interfaces, including perceived social presence, to support sustained engagement.

\textbf{Conclusion:} We demonstrate the acceptability of a voice interface to support  home practice for MBSR participants among trained mindfulness facilitators. Based on our findings, we outline design recommendations for technologies aiming to provide longitudinal support for mindfulness-based interventions. Future work should further these efforts towards making non-addictive pain management interventions accessible and efficacious for a wide audience of users.
\end{abstract}

\keywords{mindfulness-based stress reduction, voice interface, smart speaker, chronic pain}

\maketitle

\section{Introduction}

Chronic pain is a serious public health issue affecting approximately 20\% of the US adult population \cite{zelaya2020chronic}. Chronic pain is also the leading cause of disability in the United States and can result in impaired physical and mental functioning as well as reduced quality of life \cite{simon2012relieving}. Although many treatments, both pharmacological and nonpharmacological, are available for managing chronic pain \cite{skelly2020noninvasive}, an estimated 5 to 8 million Americans are prescribed opioids for long-term pain management \cite{reuben2015national}. Opioid pain medication is particularly effective in short-term pain management, but poses serious health risks since it can lead to misuse and addiction. The widespread use of opioid pain medication has contributed to a national epidemic of addiction in the US; more than 80,000 deaths occurred in 2021 alone due to opioid pain medication overdose \cite{NIDA_OD}.
As such, there is an urgent need to develop and deploy non-addictive chronic pain management methods.

Reports from the National Pain Strategy (NPS) \cite{nps2015comprehensive} and Institute of Medicine \cite{iom2011relieving} emphasize the need for evidence-based strategies that address the biopsychosocial nature of this problem. Furthermore, recent guidelines from the Center for Disease Control (CDC) on chronic pain included a recommendation on the preferred use of nonopioid treatment over opioid therapy \cite{dowell2016cdc}. These initiatives speak to the importance of advancing work on improving the accessibility of noninvasive, nonpharmacological treatment of chronic pain.

\subsection{Mindfulness-Based Strategies for Chronic Pain Management}

Toward this goal, a number of recent studies have used mindfulness-based interventions for chronic pain management \cite{toivonen2017web,veehof2016acceptance,rosenzweig2010mindfulness,cherkin2016effect,morone2016mind}. Recent studies have identified Mindfulness-Based Stress Reduction (MBSR) to be a promising alternative for long-term pain management \cite{hilton2017mindfulness}. MBSR is a particularly popular approach since it was developed specifically for patients with chronic pain who did not have their needs met by the traditional medical establishment \cite{kabat-zinn2011reflections}. MBSR was designed to increase psychological distress tolerance \cite{nila2016mindfulness}, in part, enabling individuals to disentangle the emotional and fear-based aspects of pain from the physical sensations \cite{maccoon2012validation}. 
While external sensations of pain may remain unchanged, the accompanying negative emotional and cognitive processes (e.g., hurt, suffering) of the pain experience can be reduced through this disentanglement \cite{kabat1982outpatient}. This is particularly helpful in the context of chronic pain, due to the persistent and often lifelong nature of the experience.
A significant body of work supports the efficacy of MBSR in reducing the adverse impact of chronic pain stemming from migraines \cite{wells2021effectiveness,seminowicz2020enhanced}, lower back injuries \cite{anheyer2017mindfulness,smith2020systematic,cherkin2017two}, and other conditions \cite{khoo2019comparative,rosenzweig2010mindfulness}. MBSR has also been shown to significantly reduce perception of pain intensity and functional limitations \cite{veehof2016acceptance,cherkin2016effect,morone2016mind}.

Despite these promising findings, a number of challenges can hinder the use of MBSR practices for chronic pain management. Because the efficacy of MBSR partially depends on engaging in the practices, long-term and regular home practice is essential for effective pain management, and thus reduction in opioid use. However, this can be particularly challenging for those new to mindfulness, leading to problems with treatment compliance and impacting outcomes for those engaged in the program \cite{rosenzweig2010mindfulness}. These challenges are particularly impactful for individuals living with chronic pain due to the heightened physical and emotional discomfort that can arise in the initial stages of the intervention \cite{marikar2023mindfulness}. Given the dose-response relationship between duration of MBSR practice and its degree of effectiveness, low adherence can reduce its usefulness for chronic pain management \cite{carmody2008relationships}. In other words, long-term engagement with MBSR practices is essential for effective pain management and subsequently reducing the risk of opioid dependence.

\subsection{Voice Interfaces to Support Engagement with Mindfulness Practice}

A large body of work has examined the utility of technology in supporting the practice of mindfulness \cite{parsons2019designing,li2022meditation,terzimehic2019review}. Prior work has leveraged a diverse range of technological systems, including smartphone apps \cite{lalloo2015there}, web interfaces \cite{bhattarai2017role,toivonen2017web}, and other technology-based systems (e.g., biofeedback, virtual reality) \cite{bruggeman2018hiatus,gromala2015virtual,sliwinski2017review}. While these systems have been relatively successful in supporting mindfulness outcomes, the visual and tactile interactions may present challenges for individuals with limited motor functioning \cite{anthony2013analyzing,trewin2013physical}. Given individuals with chronic pain might have limited mobility, touch or click driven interfaces can pose serious accessibility challenges for them. Furthermore, long-term engagement with these technologies tends to be quite low. 

Voice interfaces have recently emerged as a promising tool to promote user engagement with health interventions. In this work, we define ``voice interfaces'' as systems for which interactions are primarily voice-based, such that the user talks to the device and the device responds with a synthesized voice (e.g., Amazon Alexa \cite{amazon-alexa}, Google Assistant \cite{google-assistant}). Voice interfaces offer several advantages, including convenience, simplicity and confidentiality \cite{tsoli2018interactive}. A number of studies have found that such conversational technology can promote social rapport \cite{lucas2014its} and establish a trusting relationship with human participants \cite{purington2017alexa}.

A growing body of work has documented the incremental effects resulting from the use of these technologies in healthcare settings, particularly for management of chronic conditions \cite{piette2000interactive,berube2021voice}. A recent meta-analysis found that voice-driven technologies are effective in promoting adherence to health-promoting behaviors, such as medication taking and disease prevention \cite{kassavou2018automated}. Prior research concerning the application of voice interfaces in health contexts has highlighted health tracking and monitoring, assistance in locating health providers, and collecting data to aid in self-driven decision making \cite{neustein2013mobile,bafhtiar2017providing} as promising applications. Meta-analyses of customer reviews revealed that users are primarily interested in utilizing voice interface technology for self-management, as a memory aid, or for overcoming accessibility issues \cite{coyne2017early,pradhan2018accessibility}. The Amazon Alexa platform has been leveraged to assess deaf speech \cite{bigham2017deaf}, provide task support for individuals with cognitive disabilities \cite{carroll2017robin}, and increase physical activity among overweight or obese cancer survivors \cite{hassoon2018increasing}. Voice interfaces have further been shown to uniquely support positive health outcomes within vulnerable contexts, such as metastatic breast cancer \cite{qiu2021nurse,gordon2023addressing} and social anxiety \cite{wang2020alexa}. These results highlight the potential for voice interfaces to support behavior change and improve clinical outcomes, particularly for vulnerable populations. Prior work has further pointed to the utility of voice interfaces specifically for individuals living with chronic pain \cite{brewer2018accessible,fajardo2017mobile}.

While some researchers have utilized conversational technologies to support mindfulness interventions \cite{hudlicka2017enhancing,shamekhi2018breathe}, few have focused on voice interfaces. One study from Naylor et. al. \cite{naylor2008therapeutic} shows the potential for voice interfaces. Using a CBT approach, they used a phone-based interactive voice response system to support self-tracking of pain control and mood as well as provide guided assistance in practicing therapeutic skills. However, at this time, there is no existing work which uses a voice interface to delivers evidence-based mindfulness interventions specifically designed for individuals with chronic pain. While a number of mindfulness applications exist currently for the Amazon Alexa ecosystem \cite{rego2020blind}, none are oriented specifically for chronic pain management. Moreover, these applications are not informed by clinically-validated research efforts. As a result, there is no empirical evidence of their effectiveness.

\begin{figure*}[htbp]
\centering
\includegraphics[width=0.9\textwidth]{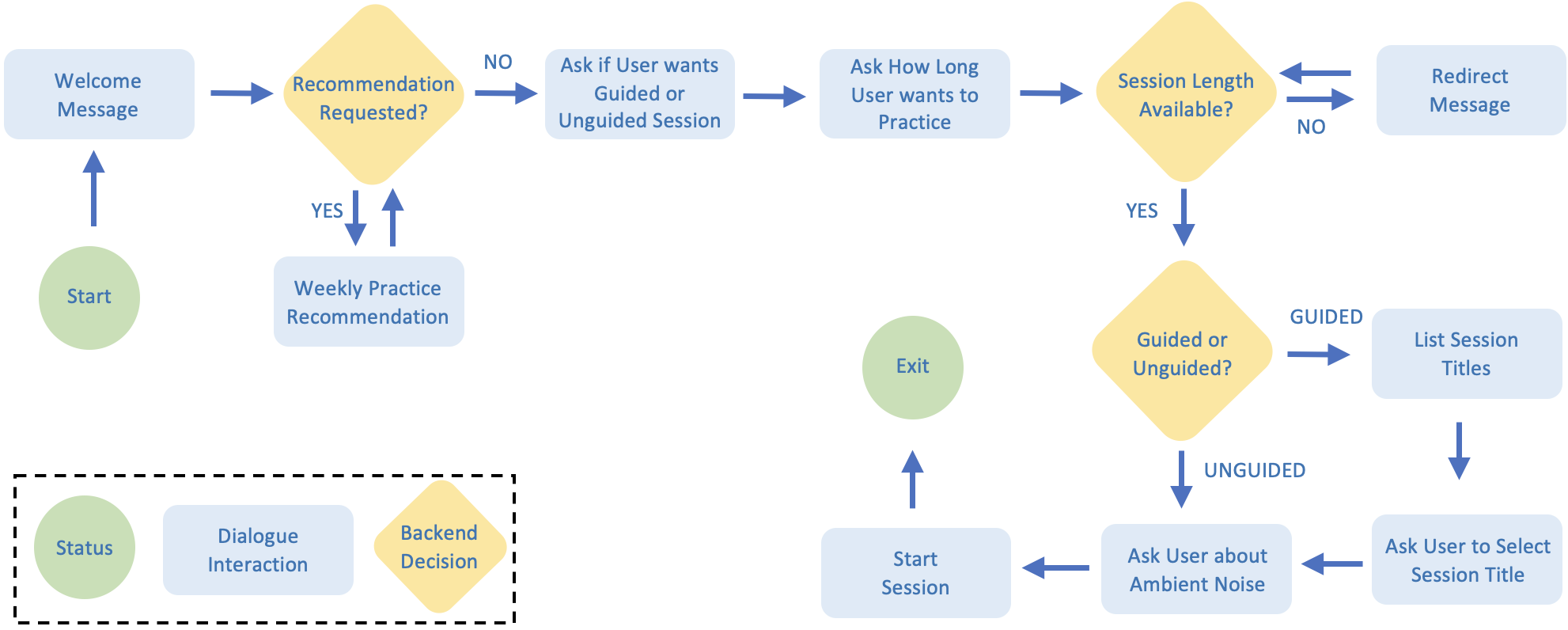}
\caption{\emph{User Flow Diagram}. The interactive dialogue structure includes branching logic based on user input and spoken utterances (i.e., dialogue interactions) implemented using the Amazon Alexa SDK \cite{alexaSDK}.}
\label{fig:user-flow}
\end{figure*}

\subsection{Current Study}

In this work, we developed a voice interface to deliver on-demand mindfulness practices. Specifically, we leveraged the Amazon Alexa ecosystem to support personalized delivery of MBSR practices designed for individuals living with chronic pain. By extending the current capabilities of Alexa, we designed a personalized and engaging virtual mindfulness practice support tool. This approach has the potential to support high adherence to MBSR practices over a long period of time due to the potential for relationship building between users and the technology. Furthermore, using an interactive voice interface may eliminate a key barrier to the success of MBSR by improving the accessibility of home practice, particularly for individuals living with chronic pain. Finally, this approach is highly scalable and can  improve access to mindfulness practices for underserved populations, including those in rural communities who face disproportionately negative consequences due to opioid prescriptions for pain management.

The goal of the current study is to understand the acceptability of this voice interface among mindfulness experts to establish the utility of this technology in supporting MBSR home practice. Specifically, we conducted semi-structured interviews with ten certified mindfulness facilitators to understand how such a technology might fit in the context of the MBSR class and identify potential usability issues with our prototype. By drawing on their rich understanding of the principles of MBSR and participant challenges informed by years of practice and experience \cite{parsons2019designing}, we have gathered a wide array of meaningful insights on how voice interfaces can effectively support mindfulness practice and what is being taught in MBSR training. Based on our findings, we have outlined a number of design recommendations for future developers of voice interfaces to support evidence-based mindfulness interventions.

\section{Method}

\subsection{System Design}

For this study, we developed a voice interface to deliver MBSR practices. Introduction to the MBSR practices was adapted to be short, interactive, and appropriate for the Amazon Alexa dialogue system. This content generation was led by the second author who completed a 6-day MBSR intensive teacher training. Requirements for participation in the teacher training included (1) one year of personal mindfulness practice, (2) participation in a previous MBSR training, and (3) participation in a 5-day silent retreat. The second author is also a certified facilitator and master trainer for two other mindfulness-based programs. She recorded all practices available on the application. 

The voice interface was designed to act as a virtual coach that could quickly and easily provide access to supporting resources for MBSR home practice, as well as foster engagement through establishing social rapport. To achieve an interactive voice-driven dialogue flow, the Alexa Skills Kit SDK \cite{alexaSDK} was used to implement the voice interface and dialogue models in the prototype.
To design the dialogue flow for the voice interface, turn-taking points were identified and branching logic was established to reflect user inputs and selected intervention strategy. The user flow diagram of the voice interface prototype is shown in Figure \ref{fig:user-flow}.

When the user starts the interaction, they first receive a friendly verbal greeting. This greeting welcomes the user to the interaction and provides a brief list of basic functions available through the skill (i.e. ``recommend a session'' based on progression through the MBSR curriculum or ``play a guided or unguided session''). If the user requests a recommendation, they are then asked to specify which week of MBSR they are currently on. Suggestions for practice for each of the eight weeks were derived from the MBSR Authorized Curriculum Guide \cite{santorelli2017mindfulness}. Based on the user's response, Alexa provides a tailored recommendation reflecting the home practice assigned for that week. Alternatively, if the user requests a guided or unguided session, they are then asked how long they would like to practice for. If the length they have requested is not available, they are subsequently informed and asked to modify their request.

Once the session length is confirmed, the available practices are specified if the user requested a guided session. Upon confirming the desired practice, the user is asked whether they would like ambient sounds to play in background during their practice. Finally, once all variable parameters have been specified, the voice interface plays the appropriate pre-recorded practice.  While novice users are expected to specify these parameters one at a time, the prototype is configured to allow experienced users to make more complex requests containing multiple parameter specifications in one command (e.g. ``I would like a 30-minute body scan without ambient noise''). This function is intended to reduce redundancy across longitudinal use and streamline access to embedded content. The co-author with MBSR expertise scripted and recorded all guided practices used in the prototype.

\subsection{Participants}

\begin{table}[hbtp]
\small
\begin{tabularx}{\linewidth}{cccCC}
\toprule
\textbf{PID} & \textbf{Gender} & \textbf{Age} & \textbf{MBSR Exp \newline (in yrs.)} & \textbf{MBSR and Related Exp \newline (in yrs.)} \\
\midrule
P1  & F & 72 & 10 & 10  \\
P2  & M & 34 & 0  & 6   \\
P3  & F & 46 & 6  & 11  \\
P4  & F & 62 & 3  & 3   \\
P5  & F & 47 & 7  & 12  \\
P6  & F & 50 & 0  & 5   \\
P7  & F & 56 & 1  & 1   \\
P8  & M & 39 & 12 & 12  \\
P9  & M & 71 & 5  & 5   \\
P10 & F & 75 & 23 & 23  \\
\bottomrule
\end{tabularx}
\caption{\emph{Participant Demographics}. All facilitators are Caucasian/White. 2 facilitators (P2, P6) were not trained in MBSR, but were trained in adjacent mindfulness programs.}
\label{tab:demographics}
\end{table}

We recruited 10 mindfulness program facilitators (7 female, 3 male; ages 34-75; see Table \ref{tab:demographics}) via e-mail and over social media through the second author's professional networks. Due to their rich understanding of the principles of MBSR and participant challenges informed by years of practice and experience \cite{parsons2019designing}, mindfulness program facilitators are an invaluable resource in informing the development of this technology. All participants had received formal training in facilitation of MBSR or related mindfulness-based interventions (e.g., Learning2Breathe \cite{broderick2021learning}) who had participated previously in MBSR. No monetary compensation was provided for participation, however, participants who did not have access to a physical Amazon Alexa device (i.e., smart speaker) were mailed a 2nd generation Amazon Echo Dot.

\subsection{Study Procedure}

Participants were asked to complete an online survey containing basic demographic questions and six open-ended questions regarding their current facilitation practices, prior experience with technologies to support mindfulness practice, and attitudes regarding the use of voice interfaces to support MBSR. These questions were based on prior work on eliciting feedback on system prototypes from domain experts \cite{argent2018clinician}. The online survey remained open to data collection for three months (from June to August 2021) and was designed to take approximately 30 minutes to complete (5 minutes per question).

Participants were directed to explore the prototype's functionality on a physical smart speaker device for as long as they wished. Participants who did not have access to a physical device were mailed an Amazon Echo Dot. Participants were asked to set up the device in a location of their choosing and install the prototype application on the device. Basic written instructions regarding the setup process were provided via email and the first author remotely provided technical support as needed. The interaction task was intentionally left unstructured (i.e., no instruction regarding tasks beyond setup) to capture the intuitiveness of the conversational flow to a naive user. Data regarding the length of interaction and specific dialogue selections was not collected. 

Following the interaction, participants were asked to evaluate the usability of the prototype via the System Usability Scale (SUS) \cite{brooke1996sus}, followed by a series of open-ended questions regarding the perceived efficacy and impact of the tool as well as suggestions for improving the current prototype. The SUS is an empirically validated scale \cite{bangor2008empirical} designed to measure users' subjective ratings of a system's usability, and is composed of 10 statements that are scored on a 5-point scale of strength of agreement (0 = ``Completely disagree'' to 4 ``Completely agree''). Final scores for the SUS can range from 0 to 100, where higher scores indicate better usability. A SUS score of 70 is generally considered to be an indicator of good usability, while a score below 50 indicates significant usability concerns \cite{bangor2009determining}. After the survey, we conducted one-on-one follow-up interviews with participants. Interviews followed a semi-structured protocol (see Supplementary Material) using the same open-ended questions from the survey as a starting point. Interviews were conducted online via Zoom and lasted approximately 60 minutes.

\subsection{Analysis}

We used directed content analysis \cite{hsieh2005three} to identify key themes and sub-themes within the interview data. Interviews were audio-recorded and transcribed for coding. We began by deriving an initial set of codes corresponding to individual questions from the semi-structured interview protocol (see Supplemental Material). The first author then manually coded facilitators' responses (i.e., open-ended survey question responses, interview transcripts) with support from the second author accordingly. The initial set of codes was iteratively expanded to accommodate emerging themes in the data which were not captured within the interview protocol. The final codebook was then circulated among members of the study team to establish consensus. Data saturation was established by observing the repetition of existing codes across different participants' transcripts, as well as the lack of emerging codes across later iterations of coding.

\section{Results}

Based on our analysis, we report findings on facilitators' current techniques for supporting home practice, their perceptions and expectations of technology to support mindfulness practice, their responses to the voice interface prototype, and potential implications for their class participants, particularly those living with chronic pain.

\subsection{Identified Needs in Current MBSR Practices} 
The need to support regular home practice was a common theme in the interviews. Facilitators described different challenges and their strategies for supporting practice outside of the formal sessions. Based on facilitators' feedback, we have highlighted identified needs and the potential of technologies to address these needs in the following section.

\subsubsection{Supporting Engagement}
The structure of MBSR requires engagement with different practices across weeks. Identifying appropriate content and practices for a given week can be confusing for MBSR participants. Facilitators reported using handouts and email reminders. However, there is a clear gap when it comes to effectively communicating how individuals should engage with MBSR practices as P8 pointed out: ``\emph{[A] lot of people do [get confused about what practices they should do], even when [we give] them a handout, even when it's written down, even with the reminder email}'' (P8). This lack of knowledge can deter individuals from engaging with regular home practices.

Facilitators also primarily used email to provide links to different MBSR resources. However, one participant noted that doing so could create additional barriers for MBSR participants to engage with in-the moment practices. P7 commented that ``\emph{any time you put any kind of barrier of `I have to search for an email that has the meditation that I'm supposed to do', that's just another thing to stop them from doing it}'' (P7).

\subsubsection{Tracking Practices and Experiences} 

The facilitators expressed interest in collecting both quantitative and qualitative measures to assess attributes of practices as well as track personal experiences. P1 noted that the quantitative data could help facilitators to address non-engagement issues: ``\emph{If you haven't done it, you know, what's getting in the way? Is it because you're so busy? Or is it because there might be a little bit of resistance?}'' (P1). P2 also commented on the usefulness of engagement data to identify potential obstacles: ``\emph{Ideally, it [will be] cool to be able to see that, on average, students practice this many times a week [\dots] Because from an instructor's perspective, you want to make sure that [\dots] they're not experiencing obstacles to keep them from practicing}'' (P2). 
There was also a consistent focus on collecting qualitative data to better understand an MBSR participant's experiences throughout the learning steps. For example, P2 commented that ``\emph{I think [it] is also important to collect [data] which is sort of around what was the meditation experience like for you? What did you notice during that experience? I think that information is really helpful too}'' (P2). 

Facilitators also expressed interest in receiving a summary of logs describing the frequency and types of practices their class participants engaged in throughout the course. P7 was particularly excited about this potential feature ``\emph{because people may be more inclined to use it if they know their information is sent to [the facilitator]}'' (P7). P8 advocated for this extension as well, arguing that it would more clearly distinguish the experience afforded by the voice interface from that of traditional offerings: ``\emph{I think you might get some more legs for innovation if you integrated performance feedback, such as it tallies up how many times they did recordings and gives that back to them}'' (P8). 

In parallel with excitement about the potential positive implications of collecting such data, facilitators emphasized the importance of ensuring that tracking does not lead to competitive comparison or counterproductive behaviors: ``\emph{I love [the] idea of building a group, [but] what I don't want to have happen is [class participants to think], 'Oh wow, they meditated a lot more than I did this week.}''' (P2). Thus, while the collection of qualitative and quantitative data to could support MBSR facilitation and participant engagement, there is a need to consider potential harms if this data is shared in a group setting.

\subsubsection{Availability and Rigor of Supporting Resources}

Facilitators talked about different forms of external resources they provided their class participants to facilitate home practice. Ideally, for MBSR, facilitators create their own set of recordings for guided home practice. However, facilitators often found it challenging to create their own recordings. Four facilitators in our study reported that they had not yet created those recordings due to time constraints and instead relied on resources published by reputable sources (e.g., Jon Kabat-Zinn, Tara Brach, Jack Kornfield). However, there were concerns about the quality of available resources. While channels in YouTube might allow easy access to these resources, the use of commercials can be interruptive to MBSR practices as P4 noted: ``\emph{I try to stay away from YouTube videos. Main reason is because [of] commercials}'' (P4). P9  questioned the general quality of online resources: ``\emph{So many recordings online are really meant to be superficial [\dots] And because a lot of people will be attracted for various reasons, [\dots] it's really important to have the depth of teaching there}'' (P9). P9 also raised concerns about the financial burden of accessing these resources due to a recent surge in monetization: ``\emph{[Online resources are] becoming monetized now [\dots] money has a way of taking precedent, and when it takes precedent, it just ruins everything}'' (P9).

Beyond publicly available online resources, facilitators also reported ``\emph{having a couple of trusted sources for these recorded meditations and [linking] clients to those resources}'' (P4). Two facilitators reported contributing to a shared repository of facilitator recordings and were thus able to leverage the full set of recordings through an institutionally crowdsourced effort. This strategy was advantageous since these facilitators' students had access to multiple options for any given practice. That said, having access to these resources does not guarantee user engagement as P2 pointed out: ``\emph{I will have a lot of people say to me, `Yeah, we have these audio files but \dots'}'' (P2). As such, there is a need for accessible and low-cost content that can support effective user engagement.

\subsubsection{Support for Community Interactions}
Facilitators noted the importance of the \textit{sangha} (community) as an important element of MBSR. P2 commented ``\emph{from a very historical perspective, there is a deep sense of community and interpersonal relationships that were really, really important to practicing meditation}'' (P2). The cultivation of trust and interpersonal relationships within the sangha of a facilitator's classroom is important for MBSR outcomes and in-class inquiry. P6 noted that, without building a foundation of trust, some MBSR participants might not feel comfortable sharing their experiences and thus may not engage in the depth of inquiry that will afford them the best outcomes.

Facilitators also pointed to the utility of community interaction in promoting accountability for practices outside of class. P3 shared that ``\emph{consistently people say it's easier to practice when you're in the group [\dots] and have that accountability}'' (P3). P7 further argued that ``\emph{If you don't have an expectation [that you should practice], you're probably not going to do whatever it is that's on your plate}'' (P7). These findings highlight the need for both the in-person sessions and home practice.

\subsection{Potential of Using Voice Interfaces to Support MBSR Practices}
This section reports on facilitators' perspectives on the potential and implications of using voice interfaces to support MBSR practices. Attitudes towards online and remote technologies in the context of mindfulness are discussed. Additionally, the unique affordances of voice interfaces for supporting MBSR home practice, such as improved accessibility, perceived social presence, and support for in-the-moment practices is considered.

\subsubsection{Affordances of Technology-Supported Mindfulness Practices}

Facilitators highlighted the utility of technology in reducing time and cost burdens related to practicing mindfulness. Technology-driven alternatives to in-person mindfulness offerings allowed for agency to practice when they wished as opposed to being restricted by class schedules. P1 expressed that the flexibility afforded by online offerings presented a unique opportunity for practices to be completed at the users' preferred time and place: ``\emph{I can [participate in yoga class] in my pajamas [\dots] Whereas with the class offerings, I had to make myself available when the class is being taught}''(P1). P3 also noted the reduced logistical burden following the adoption of remote technology: ``\emph{You decreased time stress because you cut out the commute for people to attend class [\dots] they were doing the practices in their class and then they walked out into their regular life}'' (P3). Facilitators pointed to an increasing interest from students to use technologies to support mindfulness practices as noted P1: ``\emph{I know that sometimes they would even ask for apps and which ones we might suggest. I think some of them would like it}'' (P1). 

Online technologies can be particularly useful in supporting in-the-moment practices following an individual's need and availability. P1 pointed out how such technologies can enable mindfulness practices beyond the traditional MBSR class schedules for certain groups: ``\emph{If I were a night owl, [technology] would be another helpful option for someone who might use yoga and meditation as a way to begin to relax to sleep later}'' (P1). Similarly, P7 noted how technology can enable and support in-home practices: ``\emph{If you're meditating and practicing in your home environment, absolutely, that becomes your normal. And it should because that's where you're gonna be expected to do it going forward}'' (P7).

\subsubsection{Improving Accessibility}
Facilitators highlighted that individuals living with chronic pain often have unique accessibility needs. It can be particularly challenging for them to use touch or click driven interfaces. P6 noted that ``\emph{people who experience chronic pain not only may have general mobility issues but even just using a phone, like the fine motor aspect of opening the app or whatever may be difficult}'' (P6). As a result, it can be particularly challenging for them to use touch or click driven interfaces. Voice interactions are uniquely poised to address this accessibility challenge. P5 commented that, ``\emph{for people who are in very severe pain and may be bedridden, maybe they're not able to lift their arms or wiggle their fingers [\dots] yet they can usually open their mouth and then Alexa can respond with this skill}'' (P5). Voice interfaces are uniquely poised to address this accessibility challenge

Facilitators highlighted the potential use of voice interfaces for specific vulnerable populations with limited motor function including Amyotrophic Lateral Sclerosis (ALS) . P6 described her experience with her father's ALS diagnosis and emphasized the potential of this technology for individuals with limited motor function but full capacity of verbalization. P4 also noted the implications of the voice interface for her end-of-life clients who are bedridden but can still vocalize their requests.  P9 noted how the accessibility of voice interfaces might lead to effective integration into traditional medical care: ``\emph{I think it's a great idea because I could see this easily plugged in in hospitals}'' (P9).

\subsubsection{Supporting Home Practices}

Facilitators expressed excitement about the potential for the interface to support home practice for MBSR participants. They particularly appreciated the ease of access to meditations and believed that the simplicity of asking for meditations via voice command would effectively reduce obstacles to home practice for MBSR participants. P2 commented that ``\emph{just the accessibility [barrier] stops people from practicing [\dots] It would be great if our students could just talk to Alexa and get their meditations that way because it's always so hard to get people to practice outside of class}'' (P2). P9 noted how improved accessibility can support in-the-moment practices: ``\emph{[If] somebody wants to mediate, they can meditate any time, day or night, they have the Alexa there}'' (P9). P5 also noted how accessibility of voice interface can complement current learning steps:``\emph{I think it will be really a huge difference for people taking the classes and afterward to do the daily work}'' (P5). Similarly, P6 commented: ``\emph{I think this is a really interesting way to help support home practice}'' (P6).

While facilitators agreed that the voice interface would be useful for meditation beginners, they expressed mixed views on its utility for experienced meditation practitioners. P1 speculated that voice interfaces to support MBSR practices might be mostly useful for beginners: ``\emph{If I were new, it would probably be helpful. But I've been practicing for 15 years so I don't know if I would use technology to help me}'' (P1). However, even though P5 is an experienced practitioner, she noted that guided practices delivered through the voice interface could be very useful: ``\emph{I've been meditating for [\dots] 30 years. I can really just kind of turn my brain off and let that take over and follow that [\dots] I wasn't aware of what a little gem [the guided practices] would be. And so I think that it'll be hugely beneficial for people who are well experienced with MBSR as well}'' (P5).

\subsubsection{Unique Affordances of Voice Interfaces}
Facilitators consistently pointed out the unique affordances of voice interfaces to support MBSR practices including ease of use, high interactivity, and perceived social presence. P5 commented: ``\emph{It's just so easy! [\dots] You don't have to even hardly think about it. You just have to open your mouth and there it is. It's just amazing}'' (P5). Similarly, P6 also highlighted the ease of access provided by a voice interface compared with the current methods to provide MBSR resources: ``\emph{I definitely think it's easier [than using audio files] I didn't need to have my phone or my computer open. All I needed to do was sit and voice my want of whatever type of meditation}'' (P6). Such ease of access can be a critical factor in supporting MBSR practices as P2 noted: ``\emph{People are just looking for super accessible ways. And so that's why I really like using this system. It's just always there}'' (P2).

Facilitators also noted the highly interactive nature of the voice interface, particularly its potential utility in sustaining home practices. P2 commented: ``\emph{It would be great if our students could just talk to Alexa and get their meditations that way because it's always so hard to get people to practice outside of class}'' (P2). P5 agreed that the voice interface incentivized her to practice in a way that she would not otherwise have been motivated to with other technological platforms with passive content delivery: ``\emph{If all I have to do is say ‘Hey Alexa, play a meditation' and she does, then that's great [\dots] I probably won't go through the effort of YouTube}'' (P5). Furthermore, facilitators thought the voice interface could effectively complement current MBSR teaching processes. P6 commented: ``\emph{I just am so excited about the idea of this being applied to MBSR and to facilitate home practice[\dots] We invite our participants to do this, but are we giving them enough support and direction to be able to do that on their own, and this actually helps that process, right? So it kind of closes the gap}'' (P6).

Participants also pointed to the perceived social presence of the voice interface as a positive attribute to support MBSR practices. For example, P5 commented: ``\emph{My husband travels a lot for work [\dots] and so he's just been gone a lot. And so here I am, kind of talking to Alexa and it's just been really fun to have this voice in my life [\dots] I think that just hearing her talk and being able to talk and [have] somebody [who] talks back when you speak out loud, I think that that's gonna be really, really beneficial for people trying to get through the program}'' (P5).

\subsection{Identified Design Challenges for a Voice Interface Supporting MBSR Home Practice}
We used the System Usability Scale (SUS) \cite{brooke1996sus} to assess the developed prototype. The average SUS score for the prototype voice interface was 54.8 (SD = 25.6) out of 100. More importantly, the SUS scores had a wide range with a low score of 7.5 and a high score of 97.5. These scores reflect the varying level of acceptance and expectations from the facilitators for the developed voice interface prototype. We used the interview data to get further insights into challenges and opportunities for designing a voice interface to support MBSR practices.

\subsubsection{Flexibility in Matching User Utterances and Intentions}
Facilitators consistently noted the need for flexibility in handling user utterances for different voice interactions. Not recognizing user intent due to alternative words or minor discrepancies can lead to frustration over time. P1 commented: ``\emph{One thing that frustrated me sometimes is the need to say exactly the right phrase}'' (P1). Addressing this usability issue will require moving away from rigid phrase matching to identify user intentions.

For example, P9 recommended having a list of alternative keywords that might be used to invoke different practices: ``\emph{like the auto fill function, but with voice. Now if somebody's in chronic pain, `Do that meditation thing!' And they won't remember the exact title, so there has to be like a keyword function in there. So `The meditation thing! Oh, no, no. Not that one, the one that's lying down.' To give them a lot of leeway in terms of being able to get to the meditation or the exercise that they want}'' (P9).

\subsubsection{Documentation and Support}
It is critical to provide documentation and initial support to communicate the capability and features of the voice interface. 
Three facilitators wanted written documentation outlining the structure of the interaction. P6 mentioned the importance of having step-by-step instructions specifically for older users: ``\emph{for somebody like my mom who really likes to have step-by-step directions written out for her. She's nervous she's going to mess things up, like break the technology [\dots] So having just a way to show how `If you want this, say this. If you want that, say this. If you want to exit, say this', just so [the user] understand[s] how to get where they want to go}'' (P6).

It is also necessary to provide support for initial setup of voice interfaces. P6 commented that: ``\emph{I just fear for those who really don't understand technology and don't understand apps and this and that, how much support they would need to be able to get this set up? I think once they're set up, they would be able to do it just fine. It's just the initial piece of how to work it}'' (P6). 
For four facilitators, the invocation phrase (``Alexa, open mindful pain management'') was also challenging to remember, which resulted in difficulty getting the prototype to respond. Overall, it is essential to provide both written and voice clarifications to reduce obstacle and avoid confusion.

\subsubsection{Error Handling}
Successful user interactions require graceful error handling. While a voice interface for to support MBSR home practice might not be able to address all user requests, facilitators suggested adopting flexible approach by providing closest available option to users: ``\emph{Let's say I say `3 minutes' [of meditation] but there's not a 3 minute one, so she doesn't recognize it. [In that case, she should ask:] `I have a 5-minute meditation available. Would you like to do that?' or if I said 45 minutes, which is longer than the longest one you have, ‘Oh I have a 30 minute one available. Would you like that one?'  [\dots] if it's a number that she doesn't have, routing her to the closest number}'' (P3).

It is also important to communicate rationale for not exactly matching user's initial request in these cases. Otherwise, it can lead to confusion and frustration. P10 noted feeling unsure about why her requests were not being understood by the device: ``\emph{I was asking it a question and then it was getting confusing and it was either shutting off or was it giving me something else. And so I was left feeling, as the user, that unless I had asked for whatever one in some sort of sequence which I was unaware, I wasn't gonna get it}'' (P10). This misunderstanding can be avoided by asking for user confirmation before beginning a practice and clearly communicating the system's understanding of the user's request. Although this communication may result in occasional redundancy, the resulting clarity might outweigh the potential user burden.


Facilitators also wanted a more proactive recovery from errors. Early termination of an interaction may point to potential errors and a voice interface should proactively offer related options in such cases. P9 provided an example of proactive interaction: ``\emph{`Alexa, stop! Stop it!' And then what I would recommend is Alexa says `Oh, okay. Would you like a different exercise or would you like me to stop altogether?' `No, the lying down one, the lying down.' `Okay' `I think you mean the body scan meditation. Is that correct?'}'' (P9). Such proactive error recovery can help to reduce user frustration.

\subsubsection{Options to Use Different Voices}
For a voice interface aiming to support longitudinal engagement in home practice, facilitators indicated that it may be useful to provide alternative voices. P1 suggested ``\emph{to have a variety of different voices, because all of us have preferences. Some voices we like, some we don't like, some are neutral. [\dots] So I would probably suggest having multiple options, female voice, male voice, just a variety}'' (P1). Similarly, P2 also noted the importance for different options and variety: ``\emph{It's like, `Oh, yes, I do want this voice because it's comforting, and I'm familiar with it', and other people might be like, `Well,  I've already heard this voice every week  in  a classroom setting, and so maybe I need another one'. So yeah, I think choice and variety will be something to think about}'' (P2). P1 also wanted to make alternative voices available ubiquitously available throughout the voice interface for MBSR: ``\emph{I think the [voice] option should always be there. I understand it makes more work, but I still think the option should always be there every day, not just in the beginning}'' (P1).

\subsubsection{Supporting Customizability}
Facilitators consistently noted the potential and advantage of being able to customize the voice interface. For example, P6 commented that ``\emph{you can just customize it in so many different ways which is really exciting to me}'' (P6). Specifically, there was interest in adding custom recordings to better support facilitators: ``\emph{I think a lot of MBSR instructors could potentially be interested in, when they're teaching, if they can load their recordings into appropriate slots}'' (P3).

Facilitators recommended having a wide variety of practice lengths available to optimize ability to engage in home practice. P3 also encouraged the inclusion of a practice for working through difficulty given our interest in supporting individuals living with chronic pain. P8 suggested creating a modular script for guided practice that could be started and stopped at differing intervals to curate varied lengths of sessions. Future work should look into designing voice interfaces that can empower MBSR facilitators and participants to customize home practices in a way that adheres to the philosophy while supporting personalized needs.

\subsubsection{Inclusion of Inquiry-Based Approaches}
In addition to the current practices, facilitators expressed interest in inclusion of some form of inquiry within the voice interface response. P4 suggested posing brief questions after the practice concludes such as ``\emph{How did your body feel after that?}'' and ``\emph{What did you bring up with that?}''. P10 proposed asking the user a question if they attempt to end the practice prematurely in order to disincentivize MBSR participants from taking the easy way out: ``\emph{If you're noticing that you want to stop Alexa now or you're noticing how much time you have left or you want to ask a question of how much time you have, just see if you could notice that without acting on it}'' (P10).

\section{Discussion}

Our findings highlight key design principles of voice interfaces to support mindfulness interventions. Voice interfaces are particularly well-suited to improve the accessibility of clinically-informed mindfulness practices for individuals living with chronic pain. However, there remains a number of design and implementation challenges. Specifically, effective support for MBSR using voice interfaces will require maintaining practice fidelity. In this section, recommendations are provided for future voice interface designers and developers aiming to provide mindfulness interventions for vulnerable communities, including individuals living with chronic pain.

\subsubsection{Support Data Collection but Avoid `Competition'}
Facilitators expressed the need for data collection to provide effective support for MBSR practitioners as well as sustaining engagement with MBSR practices. This aligns with findings from prior work, which highlights the diversity of approaches used by facilitators to capture home practice towards promoting adherence \cite{parsons2017home,lloyd2018utility}. Despite changes due to the recent shift to an online format, they struggled to receive meaningful participant feedback so they could refine their teaching strategies. Interaction data collected through the interface could provide insights into personalized trends, which could lead to actionable suggestions and support from facilitators \cite{allen2021making}.

However, it is critical to ensure that such tracking is not shared in a way that promotes comparison or competition. Recent work has identified how behavioral tracking can lead to unintended negative consequences \cite{eikey2021selfreflection}. To avoid problems, designers should allow for facilitators to decide how and when to provide quantitative metrics of home practice to MBSR participants. By delegating this responsibility to the facilitator, the technology will better support rather than detract from the facilitator's personal teaching philosophy \cite{parsons2019designing}. Furthermore, the technology can thus draw upon facilitators' expertise and extensive knowledge of the MBSR curriculum without pre-determining the appropriate strategy from a limited sample of perspectives.

\subsubsection{Support Facilitator-Participant Interaction}
Additionally, extending the accountability engendered by the classroom community emerged as an important factor underlying consistent home practice. This is particularly important given community interaction is absent when MBSR participants practice at home, which can lead lack of engagement over time \cite{allen2009participants}. Efforts to support direct and accessible pathways for open communication between facilitators and their class participants are necessary to promote the acceptability of this technology in this context \cite{parsons2019designing}. As such, future work should consider supporting interactions by which MBSR participants can engage with each other and the facilitator outside of sessions. For example, knowing when someone else in their class is engaging in a mindfulness practice may encourage them to engage as well.

\subsubsection{Ensure Ease of Access}
The ease of access is a critical factor in ensuring engagement with MBSR practices. Facilitators repeatedly noted how their class participants often forgot which practices were assigned for home practice which caused them to forgo practicing altogether. These findings are consistent with prior work which point to role of both perceived (e.g., cognitive dissonance, forgetfulness) and pragmatic barriers (e.g., lack of time, physical discomfort) in reducing the amount of time individuals practice at home \cite{barcelo2023systematic,chaskalson2011mindful}. Facilitators also observed that MBSR participants were reluctant to practice at home if supporting resources were not readily available. A voice interface can effectively address these barriers by streamlining access to different practices, thus supporting in-the-moment practices and potentially contributing to sustained engagement with MBSR \cite{mclean2019hey}. Designing effective reminders and notifications through the voice interface could further mitigate existing barriers to engagement with home practice and improve the overall MBSR participant experience.

\subsubsection{Foster Social Presence}
A voice interface aiming to sustain user engagement should focus on fostering social presence. Voice interactions provide a unique affordance, when it comes to perceived social presence \cite{nass2005wired}. Facilitators also noted how voice interactions can lead to perceived social presence, which could subsequently improve engagement \cite{mclean2019hey}.
For individuals with chronic pain, this aspect of voice interface technology is critically important since physical pain may inhibit their ability to engage with others during flare ups \cite{barcelo2023systematic}. Future work in this domain should explore how voice interfaces can support MBSR practices as well as improving perceived social presence through different interactions.

\subsubsection{Provide User Guide and Documentation}
Designers should provide adequate user guide and documentation to ensure successful interactions with the voice interface. Specifically, it is critical to convey utterances and phrases that might be used to invoke different functionalities of the voice interface as it may be challenging for users to discover features and capabilities of a voice interface without explicit guidance \cite{chen2006designing,yankelovich1996users}. The complete dialogue flow should succinctly and thoroughly be communicated to the user in advance of any interactions. This can be achieved via written documentation in conjunction with verbal explanation within the voice interface \cite{corbett2016can}. While verbal explanation should be provided without an explicit request the first time the user interacts with the voice interface, users should be able to request the verbal explanation at any time via voice command. The written documentation should serve as an adjunct to the in-situ verbal explanation and provide a complete picture of every possible interaction scenario. It should further be organized such that the user can readily look up and identify their concerns, rather than being a necessary precursor to interacting with the voice interface.

\subsubsection{Handle Errors Gracefully}
Facilitators noted that individuals struggling with physical discomfort and subsequent lower cognitive functioning (e.g., individuals with chronic pain) may experience difficulty remembering voice commands and phrases, particularly in moments with intense discomfort \cite{noyes1992speech}. It is thus critical to be flexible in handling user inputs. Rigorous testing must be conducted to identify variations in phrasing (e.g. ``standing mindful movement'', ``standing movement'', ``standing mindfulness'') to avoid errors when a user deviates slightly from the expected response. In other words, the voice interface should be able to recover from most errors and identify user intents effectively. In case of unavoidable errors, the voice interface must clearly articulate the reason as well as suggest relevant options to users. Prior work supports the efficacy of these strategies in preventing negative user experiences due to command misrecognition in voice interfaces \cite{murray1993data}. Handling errors gracefully and robustly is critical to facilitate positive interactions with individuals during their moments of needs.

\subsubsection{Provide Different Voices}
As a standard feature, the voice interface should support different voices to meet individual preferences (e.g., voices representing different genders and personalities). Allowing users to customize speech characteristics of voice interfaces has been shown to promote trust, attraction, and favorability \cite{snyder2023busting, zargham2022want}. Facilitators pointed out the utility of having a diverse set of voices to choose from when offering supportive resources to their class participants. Because these preferences might vary for different practices, a voice interface should allow users to pick their preferred voice for any given interactions. Both Alexa and Google provide different voices and personas. For example, both platforms allow users to choose between a masculine or feminine voice options, with Google offering multiple within-gender variations of vocal characteristics \cite{seifert2022how,google2018new}. Prior work highlights the impact of these gender differences on users' perceptions of voice interfaces, particularly in health-related contexts \cite{goodman2023say}. Future work should look into how the use of different voice attributes can address user needs across diverse contexts.

\subsubsection{Enable Customization}
Consistent with prior work \cite{parsons2019designing}, facilitators repeatedly noted the need for editing and tailoring different activities and interactions. To adequately address these needs, a voice interface should enable facilitators to add and adapt MBSR content as necessary. Furthermore, it should support the facilitators in uploading and using their own personal guided audio-recordings of practices for their class participants. This recommendation aligns with guidelines for MBSR facilitator training, which explicitly encourages the development of individualized guided recordings by each facilitator for their own instruction purposes \cite{woods2021mindfulness}. This would require supporting modular script and audio recording through the voice interface. Furthermore, interactions and content should be adaptive to support personalized needs of a class participant. 

A voice interface should also be able to remember user preferences and prior activities to offer easy customization for subsequent sessions. While the value of persistent memory across user interactions has been established in the context of human-robot interaction \cite{baxter2014pervasive}, limited work has focused on its affordances for voice interfaces. For example, memory of a user's activities would allow the system to automatically understand the progress of an individual through the MBSR curriculum and suggest relevant practices accordingly. Future work should explore the impact of accounting for user preferences and historical interactions on the acceptability and efficacy of voice interfaces for supporting MBSR home practice.

\subsection{Limitations \& Future Work}
The study has a number of limitations. First, interview data was collected from a small number of trained mindfulness facilitators. While this sample size is consistent with prior work that also used one-on-one semi-structured interviews with stakeholders to understand perceptions and requirements of technology within a healthcare setting \cite{argent2018clinician,parsons2019designing}, future work should explore data from a larger sample of experts and supplement interview discussions with quantitative measures of usability (e.g., backend usage logs).

Additionally, these findings are limited to the perspective of facilitators which may not reflect the intended user (i.e., individuals living with chronic pain). While this approach provides useful insights towards integrating our prototype into existing MBSR program facilitation, it is critically important that future studies collect interaction and acceptance data from individuals living with chronic pain. Future work should also investigate the role of healthcare providers in facilitating the use of mindfulness practices taught in MBSR by integrating digital health tools and technologies into existing clinical practices. 

Finally, our prototype was adapted specifically to the MBSR curriculum due to the credentials and training of the research team and accessibility of relevant content experts. As such, our findings do not generalize to other mindfulness-based interventions. However, the flexible structure of our prototype could easily be adapted to support other mindfulness-based interventions (e.g., Mindfulness-Based Cognitive Therapy, Acceptance and Commitment Therapy). Future work should investigate the utility of voice interfaces in the context of other mindfulness-based interventions.

\section{Conclusion}
With the rising need for non-addictive pain management methods to address chronic pain, efforts to support longitudinal engagement with interventions like MBSR are critically important. In response to this challenge, we developed a voice interface to facilitate MBSR home practice. Findings from our preliminary evaluation show that facilitators supported the use of the voice interface for MBSR practices, particularly for individuals with limited motor function. Facilitators also highlighted unique affordances of voice interfaces, including perceived social presence, to support sustained engagement. Based on findings, we have outlined design recommendations for technologies aiming to provide longitudinal support for mindfulness-based interventions. Future development should further these efforts towards making non-addictive pain management interventions accessible and efficacious for a wide audience of users.

\declaration{Conflicting Interests}
The author(s) declare that there is no conflict of interest.

\declaration{Funding}
This work was not funded by any ongoing grants.

\declaration{Ethical Approval}
All procedures performed in studies involving human participants were in accordance with the ethical standards of the local institutional review board. The requirement for written informed consent was waived due to the minimal harms posed by the study procedures.

\declaration{Guarantor}
SM

\declaration{Contributorship}
SA, SDF, and STL researched the literature and conceived the study. SDF was involved in participant recruitment, along with SM. SM led the data collection, including conducting participant interviews. SM and SDF performed the data analysis and SA helped with interpretation. SM wrote the first draft of the manuscript. All authors reviewed and edited the manuscript and approved the final version of the manuscript.

\declaration{Acknowledgements}
We would like to thank Ryan O'Neill for his assistance in developing the initial prototype for this work.

\bibliographystyle{SageV}
\bibliography{references}

\end{document}